\documentclass[aps,prl,preprint,showpacs,groupedaddress]{revtex4}
\usepackage{graphicx}
\usepackage{textcomp}
\usepackage{amsmath}
\usepackage{amssymb}

\begin{document}
\draft
\title{Geometric potential and transport in photonic topological crystals}
\normalsize

\author{Alexander Szameit$^{1\ast}$, Felix Dreisow$^2$, Matthias Heinrich$^2$, Robert Keil$^2$, Stefan Nolte$^2$, Andreas T\"{u}nnermann$^2$, Stefano Longhi$^3$}
\address{$^1$Physics Department and Solid State Institute, Technion, 32000 Haifa, Israel}
\address{$^2$Institute of Applied Physics, Friedrich-Schiller-Universit\"{a}t, Max-Wien-Platz 1, 07743 Jena, Germany}
\address{$^3$Dipartimento di Fisica, Politecnico di Milano, Piazza Leonardo da Vinci 32, 20133 Milan, Italy}


%
\bigskip
\begin{abstract}
We report on the experimental realization of an optical analogue of
a quantum geometric potential for light wave packets constrained on
thin dielectric guiding layers fabricated in silica by the
femtosecond laser writing technology. We further demonstrate the
optical version of a topological crystal, with the observation of
Bloch oscillations and Zener tunneling of purely geometric nature.
\end{abstract}



\maketitle

Studying the quantum mechanics of a particle constrained on a curved
space \cite{DeWitt57,Jensen71,daCosta81} has been a matter of great
controversies for more than 50 years. For a long time, it has been
known that geometry influences in different ways the motion of
classical and quantum particles confined on a curved surface or on a
line. When surface curvatures become comparable to the de Broglie
wavelength, geometrical effects come into play by the internal
metric of the constraint manifold as well as by the external metric
of the embedding space. As opposed to a classical particle, a
quantum particle retains some knowledge of the surrounding
three-dimensional space and, in spite of the absence of
interactions, it experiences an effective frictional potential of
geometric nature \cite{Jensen71,daCosta81}. Such a geometric
potential is of major importance in the understanding and control of
the physical properties of novel low-dimensional functional
materials, like curved carbon nanotubes and DNA wires
\cite{carbonDNA}. For instance, an electron confined to a
periodically curved surface senses a periodic frictional potential
which acts as a topological crystal \cite{Aoki01}. The definition of
a geometric potential in quantum and condensed matter physics,
however, is troubled by the realization of a proper squeezing
procedure \cite{Kampen84,Kaplan97}, which is needed to avoid
operator-ordering ambiguities \cite{Schulman81,Ikegami92}. A rather
accepted approach that avoids operator-ordering ambiguities is the
confining-potential approach, originally proposed by Jensen, Koppe
and da Costa (JKC) \cite{Jensen71,daCosta81}, in which a strong
force acting normal to the surface provides the appropriate
confinement. Quantum excitation energies in the normal direction are
raised far beyond those in the tangential direction. Hence, the
particle motion normal to the surface can be safely neglected, which
leads to an effective Hamiltonian for propagation along the curved
surface. Though the JKC approach has gained a broad consensus
especially in the theoretical condensed matter physics community
\cite{Entin01,Gravesen05,Zhang07,Ferrari08,Shima09}, it was noticed
that such an ideal squeezing procedure may correspond to unrealistic
restrictions or cannot be unique \cite{Kampen84,Kaplan97}. Clear
evidences of geometric potentials in low-dimensional curved
nanosystems are still lacking and are likely to be a controversial
matter. On the other hand, optics has offered in recent years a
fascinating laboratory tool to investigate classical analogues of
otherwise inaccessible quantum-mechanical and relativistic effects
(see, e.g., \cite{Schwartz07,Philbin08,Longhi09,Szameit09,Genov09}).
 In this Letter we report on the first experimental realization
of an optical analogue of a quantum geometric potential for light
wave packets constrained on thin dielectric guiding layers
fabricated in silica by the femtosecond laser writing technology
\cite{Itoh06}. We further demonstrate the optical version of a
topological crystal \cite{Aoki01}, with the observation of Bloch
oscillations and Zener tunneling of purely geometric nature
\cite{Longhi07}.
\par
Let us consider wave propagation at wavelength $\lambda$ in a thin
and weakly guiding dielectric layer, which is assumed to be
invariant along the $z$ direction and arbitrarily curved in the
transverse $(x,y)$ plane along a curve $\gamma$ (see Fig.1a). In the
scalar approximation, the electric field $E(x,y,z)$ satisfies the
Helmholtz equation
\begin{equation}
\nabla_{t}^2E+\partial^2_z E+(n/\lambdabar)^2E=0,
\end{equation}
where $\nabla_{t}^2=\partial^2_x+\partial^2_y$ is the transverse
Laplacian, $\lambdabar=\lambda/(2 \pi)$ is the reduced wavelength,
and $n=n(x,y)$ is the refractive index profile of the optical
structure, which is assumed to be independent of $z$ and weakly
deviating from the refractive index $n_s$ of the substrate. The
analogy between spatial propagation of light waves along the $z$
direction and the temporal dynamics of a two-dimensional quantum
particle constrained on the curved line $\gamma$, discussed by da
Costa \cite{daCosta81}, is at best captured by the introduction of
the local curvilinear coordinates $(\sigma,\eta)$ of Fig.1a. The
volume element in the curvilinear system is given by $dV=\chi
d\sigma d \eta dz$, where $\chi(\sigma,\eta)=1+\eta/R$ and
$R=R(\sigma)$ is the local radius of curvature of $\gamma$. To study
the behavior of light waves near the guiding layer ($\eta
\rightarrow 0$), following \cite{daCosta81} it is worth introducing
the new wave field $\mathcal{E}=E \sqrt{\chi}$. After writing in
Eq.(1) the transverse Laplacian $\nabla_t^2$ in curvilinear
coordinates and with the substitution $E=\mathcal{E} / \sqrt{\chi}$,
the evolution equation of the field $\mathcal{E}$ in the limit $
\eta \rightarrow 0$ (i.e., close to the line $\gamma$) reads
\begin{equation}
(\partial^2_{\sigma}+\partial^2_{\eta}+\partial^2_{z})\mathcal{E}+
(1/4R^2)\mathcal{E}+(n / \lambdabar)^2 \mathcal{E}=0.
\end{equation}
In the weak guiding limit $|n(\sigma,\eta)-n_s| \ll n_s$ and
assuming $|R| \gg \lambdabar$, the paraxial approximation can be
introduced in the usual way by setting
$\mathcal{E}(\sigma,\eta,z)=\psi(\sigma,\eta,z) \exp(i n_s z/
\lambdabar)$ and neglecting $\partial^2_z \psi$ as compared to $(1/
\lambdabar) \partial_z \psi$. This yields the optical
Schr\"{o}dinger equation
\begin{equation} \label{eq1}
i \lambdabar \frac{\partial \psi}{\partial
z}=-\frac{\lambdabar^2}{2n_s} \left( \frac{\partial^2}{\partial
\sigma^2} + \frac{\partial^2}{\partial \eta^2} \right) \psi + \left[
V_c({\sigma,\eta}) + V_g(\sigma)\right] \psi,
\end{equation}
where $V_c(\sigma,\eta)= [n_s^2-n^2(\sigma,\eta)]/(2 n_s) \simeq
n_s-n(\sigma,\eta)$ is the {\em confining potential} and
\begin{equation} \label{eq2}
V_g(\sigma)=-\frac{\lambdabar^2}{8n_s R^2(\sigma)}
\end{equation}
is the so-called {\em geometric potential} \cite{daCosta81}. The
correspondence of Eq.(3) to the quantum-mechanical Schr\"{o}dinger
equation of a two-dimensional particle constrained on the curve
$\gamma$, discussed by da Costa \cite{daCosta81}, is formally
obtained after replacing the photon wavelength $\lambda$ with the
Planck constant $h$, the refractive index $n_s$ with the particle
mass $m$ and the spatial coordinate $z$ with time $t$. Hence, in
optics the quantum-mechanical evolution in time of a two-dimensional
wave function is mapped onto the propagation of an optical wave
packet along the spatial $z$-direction. The $z$-independence of the
path $\gamma$ reflects the circumstance that the constraint is
time-independent. Similarly to the quantum mechanical problem
\cite{daCosta81}, in the optical Schr\"{o}dinger equation (3) the
confining potential $V_c$, squeezing the wave packet around
$\gamma$, originates from the {\it physical} change of the
refractive index in the guiding layer (see Fig.1b); on the contrary,
the geometric potential $V_g$ is of {\it geometric} nature and
arises from the diffraction operator $-(\lambdabar^2/2n_s)
\nabla^2_{t}$ (the analogue of the kinetic energy operator) in
curvilinear coordinates. In the curved reference frame
$(\sigma,\eta)$, this potential acts on light waves like a
fictitious refractive index change.\\
In the JKC approach, the confining potential $V_c$ is taken to be
independent of $\sigma$, a condition which ensures that the
constraint is frictionless in the classical (geometric-optic) limit.
In this case, the motion in the normal $\eta$ direction can be
exactly separated from the dynamics by letting
$\psi(\sigma,\eta,z)=F(\sigma,z) g(\eta) \exp(-i E_0 z /
\lambdabar)$, where $g(\eta)$ is the ground-state wave function of
the confining potential $V_c$ and $E_0$ its corresponding energy
[i.e., $-(\lambdabar^2/2n_s) (d^2g/d \eta^2)+V_c(\eta) g(\eta)=E_0
g(\eta)$]. One then obtaines an effective equation for the particle
motion along the $\sigma$ coordinate
\begin{equation} i \lambdabar \frac{\partial
F}{\partial z}=-\frac{\lambdabar^2}{2n_s} \frac{\partial^2
F}{\partial \sigma^2}+ V_g(\sigma) F.
\end{equation}
Therefore, the geometric potential $V_g$ acts as an effective
frictional potential for the motion on the curved manifold.
Unfortunately, there is no compelling reason that actual confining
potentials have this exceptional property \cite{Kampen84}. Moreover,
it was shown that one can construct {\it ad-hoc} confining
potentials leading to an arbitrary correction to the geometric
potential that obey the frictionless condition of the constraint in
the classical limit $\lambda \rightarrow 0$ \cite{Kaplan97}.
Non-ideal squeezing potentials will generally replace the geometric
potential $V_g$ in Eq.(5) by an effective frictional potential
$V_{eff}$. If we allow the confining potential $V_c$ to slowly vary
with $\sigma$, the tangential motion can still be approximately
separated from the normal one by using a multiple scale asymptotic
analysis (as in \cite{Kaplan97}). In this way, the tangential motion
turns out to be governed again by the reduced Schr\"{o}dinger
equation (5), but with the geometric potential $V_g$ replaced by
$V_{eff}=V_g+\Delta V_g$, where
\begin{equation}
\Delta V_g(\sigma)=\Delta E (\sigma)-\frac{\lambdabar^2}{2n_s} \int
d \eta g \frac{\partial^2 g}{\partial \sigma^2}.
\end{equation}
In the above equation, $g(\sigma,\eta)$ is the local ground-state
wave function of the confining potential $V_c$, normalized such that
$\int d \eta g^2=1$, and $E_0+\Delta E(\sigma)$ is the corresponding
$\sigma$-dependent energy, i.e. $-(\lambdabar^2/ 2n_s)
\partial^2_{\eta}g +V_cg=(E_0+\Delta E)g$.
The JKC geometric potential is attained whenever $\Delta
V_g(\sigma)$ vanishes, a condition which {\em does not necessarily
imply} the $\sigma$-invariance of $V_c$.
\par
\begin{figure}[htbp]
  \includegraphics[width=13cm]{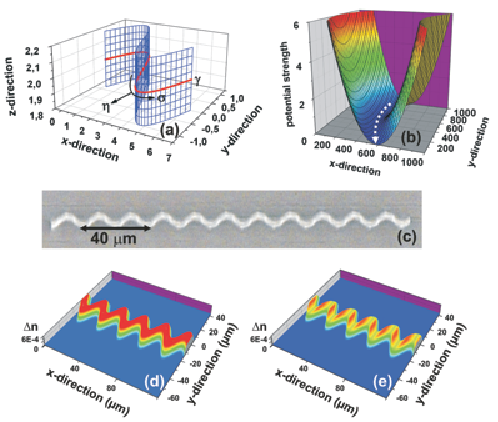}\\
   \caption{(a) Sketch of a two-dimensional cylindrical (i.e. $z$-invariant) curved surface with the curvilinear coordinates
   $(\sigma,\eta)$. (b) The forced path of a wave packet in a constraining potential. (c) Microscope image of the front facet
   of the undulating guiding layer that realizes a one-dimensional topological crystal. (d) Homogeneous refractive index change yielding
   non-ideal squeezing. (e) Corrected refractive
   index distribution, resulting in ideal JKC conditions.}
   \label{fig1}
\end{figure}

 In order to demonstrate the features of an ideal JKC squeezing
procedure and the discrepancies of a non-ideal squeezing, we
realized an optical analog of a topological crystal \cite{Aoki01},
in which a two-dimensional wave function is squeezed onto an
undulating curve $\gamma$. To this aim, a sinusoidally undulated
slab waveguide (Fig. \ref{fig1}c), defined by $y(x)=A \sin (2 \pi x/
\Lambda)$, was fabricated using the laser direct-writing technology
\cite{Itoh06}. The use of this technique commonly yields a layer
which exhibits constant thickness in the vertical $y$-direction
rather then in the normal $\eta$-direction. For a homogeneous
refractive index change as shown in Fig. \ref{fig1}d, this implies
that the correction to the geometric potential does not vanish, i.e.
$\Delta V_g \not = 0$. However, using the laser direct-writing
approach, the refractive index variations along $\gamma$ can be
tailored with great accuracy. In particular, it is possible to
modulate the strength of $n(\sigma,\eta)$ along $\sigma$ such that
$\Delta V_g$ gets negligible as compared to $V_g$. The refractive
index distribution that closely approximates the JKC condition
$\Delta V_g=0$ is shown in Fig. \ref{fig1}e, and corresponds to the
highest index change at the inflection points of the undulation,
where $R=\infty$.

\par
\begin{figure}[htbp]
  \includegraphics[width=13cm]{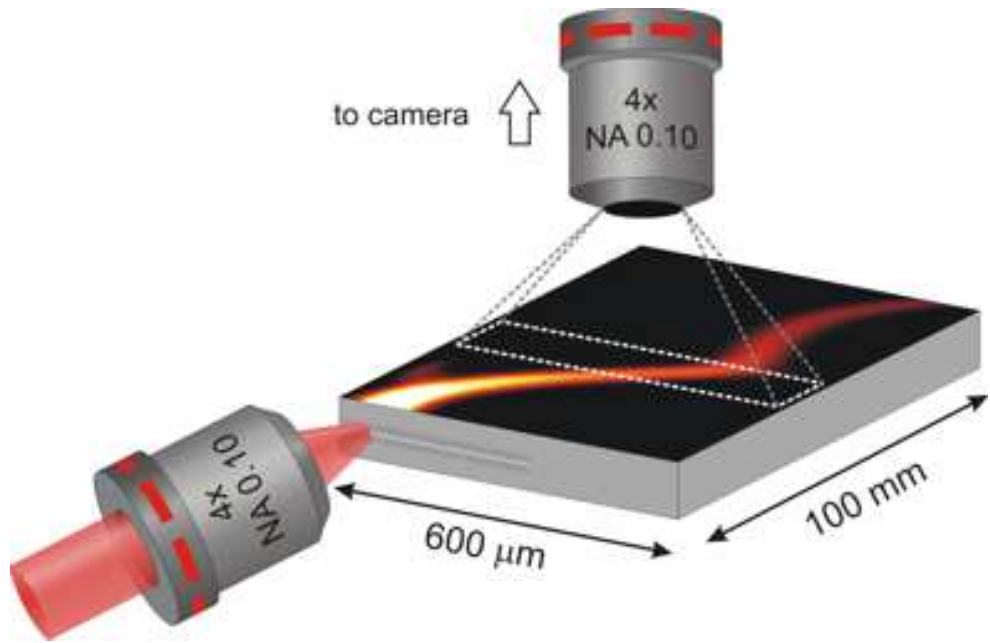}\\
   \caption{Measurement setup using fluorescence microscopy.}
   \label{fig2}
\end{figure}

\begin{figure}[htbp]
  \includegraphics[width=13cm]{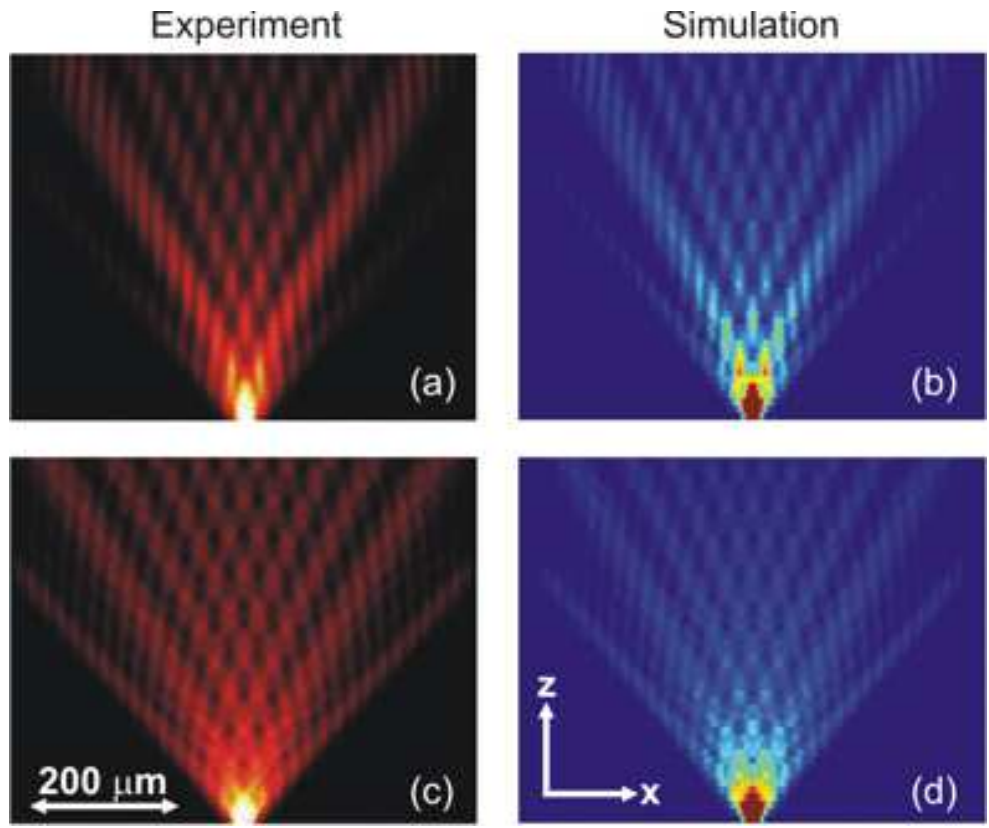}\\
   \caption{(a) Measurement of the light propagation in a non-ideal squeezing potential, and (b) corresponding numerical confirmation.
   (c) Measurement of the light propagation under ideal JKC conditions, and (d) corresponding numerical confirmation.}
   \label{fig3}
\end{figure}

  To investigate wave packet dynamics in  the topological
crystal, we employ a fluorescence microscopy technique
\cite{Szameit07}. Light around $\lambda = 633 \; \mathrm{nm}$
excites color centers, which are formed during the waveguide
fabrication process. The resulting fluorescence is proportional to
the intensity of the propagating light, and can be observed from
above the sample, as sketched in Fig. \ref{fig2}, due to the
isotropic emission. When launching a beam into the two samples with
the refractive index profiles representing the non-ideal (Fig.
\ref{fig1}d) and the ideal (JKC) case (Fig. \ref{fig1}e), one
obtains the intensity distributions shown in Fig. \ref{fig3}a,c,
respectively. The observations are confirmed by numerical
simulations of the full wave equation (see Fig. \ref{fig3}b,d).
Although in both cases the initial wave packet spreads, the
spreading rate is larger for the JKC potential. Most importantly,
the spreading pattern observed in Fig.3c reproduces with excellent
accuracy the diffraction  pattern in a one-dimensional crystal with
the potential defined by Eq.(4). Hence, our experimental technique
enables to squeeze a wave packet mimicking the ideal JKC method. It
is important to point out that the equivalent lattice potential
$V_g$ caused by the curvature of the JKC guide has a geometric
origin and cannot be naively explained by effective-index or
variational methods generally adopted in guided-wave optics
\cite{Benson}. As the effective index method well explains the
formation of periodic optical potentials in common waveguide array
settings (such as those investigated in \cite{Christodoulides03}),
it fails to explain the onset of the geometric potential for the JKC
undulating slab structure of Fig.1e. In fact, if one calculates from
Eq.(1) the effective index along the vertical $y$ direction using a
separation variable method \cite{Benson}, the resulting effective
index $n_e(x)$ \cite{note} turns out to be largest in the regions
with vanishing curvature ($R=\infty)$, where the local refractive
index is highest (see Fig.1e). This would erroneously predict light
confinement in regions around $R=\infty$, rather than where $|R|$ is
minimal [see Eq.(4)].

\par
\begin{figure}[htbp]
  \includegraphics[width=13cm]{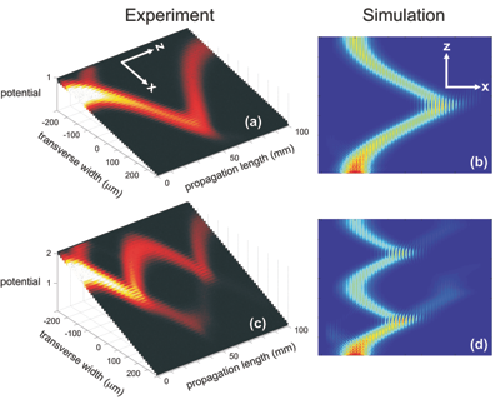}\\
   \caption{(a) Dynamics of light in an optical topological crystal for broad beam excitation, when
   a transverse dc force is applied. (b) Numerical confirmation. (c) Zener tunneling for broad
   beam excitation in a high potential gradient. (d) Numerical confirmation. Note that the transverse dc force
   is applied by slightly curving the guiding layer in the $z$-direction. For reasons of visibility,
   this curvature has been numerically removed from the experimental images. Instead, the transverse gradient is sketched by tilting the images.}
   \label{fig4}
\end{figure}

 Using the confirmed JKC potential, we finally investigated
the transport properties in the topological crystal by the
application of a direct current (dc) force. Similarly to ordinary
waveguide arrays, the dc force is expected to inhibit wave packet
spreading and to cause an oscillatory motion of the  wave packet
[Bloch oscillations (BOs)] via the formation of a Wannier-Stark
ladder spectrum \cite{Christodoulides03}, with the occurrence of
Zener tunneling (ZT) at high dc forcing \cite{Trompeter06}.
According to \cite{Lenz99}, a fictitious dc force is realized by
fabricating the undulated slab slightly curved in the $z$-direction,
which approximately preserves the $z$-invariance of the path
$\gamma$, but imprints a transverse dc force in the $x$-direction.
Our structure thus realizes a test bed for the observation of
so-called topological BOs predicted in \cite{Longhi07}, where both
effective lattice potential and dc force arise from the geometric
deformation of a slab waveguide. The experimental observation of BOs
in topological photonic crystals, corresponding to a broad input
beam excitation, is shown in Fig. \ref{fig4}a, together with the
numerical prediction (Fig. \ref{fig4}b). In order to visualize ZT,
we doubled the transverse force by halving the longitudinal radius
of curvature of the undulated slab waveguide. The results
corresponding to broad beam excitation at the input plane are shown
in Figs. \ref{fig4}c,d. Note that, as compared to Figs. \ref{fig4}a
and \ref{fig4}b, ZT is now clearly visible because a considerable
fraction of the wave packet does not experience Bragg reflection,
but tunnels into higher propagation bands.

\par In conclusion, we experimentally observed a geometric potential
for optical wave packets constrained on curved surfaces and showed
the impact of the squeezing procedure on the resulting frictional
potential. In particular, an optical version of a topological
crystal \cite{Aoki01} has been realized, in which the band structure
of the crystal is determined by the geometric potential according to
the JKC theory \cite{daCosta81}. Our experiments shed new light onto
the old and rather controversial problem of wave mechanics of
quantum particles constrained on a curved space, and open the
possibility to explore the interplay between topology and transport
in low-dimensional curved structures. Our findings could also pave
the way towards the investigation of photonic materials with
topologically-controlled diffractive and refractive properties.

\par The authors acknowledge support by the Deutsche
Forschungsgemeinschaft (Research Unit 532 and Leibniz program), the
German Academy of Science Leopoldina (grant LPDS 2009-13), and the
italian MIUR (PRIN 2008 project).

\clearpage
\clearpage
\clearpage
\clearpage


\begin{thebibliography}{31}

\bibitem{DeWitt57}
B.S. DeWitt, Rev. Mod. Phys. {\bf 29},
377 (1957).

\bibitem{Jensen71}
H. Jensen and H. Koppe, Ann.
Phys. (N.Y.) {\bf 63}, 586 (1971).

\bibitem{daCosta81}
R.C.T. da Costa, Phys. Rev. A {\bf
23}, 1982 (1981).

\bibitem{carbonDNA}
M. Zheng, A. Jagota, E.D. Semke, B.A. Diner, R.S. Mclean, S.R.
Lustig, R.E. Richardson, and N.G. Tassi, Nature Materials {\bf 2}, 338
(2003).

\bibitem{Aoki01}
H. Aoki, M. Koshino, D. Takeda, and H. Morise, Phys. Rev.
B {\bf 65}, 035102 (2001).

\bibitem{Kampen84}
N.G. van Kampen and J.J. Lodder, Am. J. Phys. {\bf 52},
419 (1984).

\bibitem{Kaplan97}
L. Kaplan, N.T. Maitra, and E.J. Heller, Phys. Rev. A {\bf 56}, 2592 (1997).

\bibitem{Schulman81}
L.S. Schulman, Techniques and Applications of Path Integration,
(Wiley, New York, 1981).

\bibitem{Ikegami92}
M. Ikegami, Y. Nagaoka, S. Takagi, and T. Tanzawa, Prog. Theor. Phys. {\bf 88}, 229
(1992).

\bibitem{Entin01}
M.V. Entin and L.I. Magarill, Phys. Rev. B {\bf 64}, 085330 (2001).

\bibitem{Gravesen05}
J. Gravesen and M. Willatzen, Phys. Rev. A {\bf 72}, 032108 (2005).

\bibitem{Zhang07}
E. Zhang, S. Zhang, and Q. Wang, Phys. Rev. B {\bf 75}, 085308
(2007).

\bibitem{Ferrari08}
G. Ferrari and G. Cuoghi, Phys. Rev. Lett. {\bf 100}, 230403 (2008).

\bibitem{Shima09}
H. Shima, H. Yoshioka, and J. Onoe, Phys. Rev. B {\bf
79}, 201401(R) (2009).

\bibitem{Schwartz07}
T. Schwartz, G. Bartal, S. Fishman, and M. Segev, Nature {\bf 446}, 52 (2007).

\bibitem{Philbin08}
T. G. Philbin, C. Kuklewicz, S. Robertson, S. Hill, F. K\"{o}nig,
and U. Leonhardt, Science {\bf 319}, 1367 (2008).

\bibitem{Szameit09}
A. Szameit, I.L. Garanovich, M. Heinrich, A.A. Sukhorukov, F.
Dreisow, T. Pertsch, S. Nolte, A. T\"{u}nnermann, and Y.S. Kivshar,
Nature Phys. {\bf 5}, 271 (2009).

\bibitem{Genov09}
D. A. Genov, S. Zhang, and X. Zhang, Nature Phys. {\bf 5}, 687 (2009).

\bibitem{Longhi09}
S. Longhi, Laser \& Photon. Rev. {\bf 3}, 243 (2009).

\bibitem{Itoh06} K. Itoh, W. Watanabe, S. Nolte, and C. Schaffer,
MRS Bulletin \textbf{31}, 620 (2006).

\bibitem{Longhi07}
S. Longhi, Opt. Lett. {\bf 32}, 2647 (2007).

\bibitem{Szameit07} A. Szameit, F. Dreisow, H. Hartung, S. Nolte,
and A. T\"unnermann, Appl. Phys. Lett. \textbf{90}, 241113 (2007).

\bibitem{Benson}
T. M. Benson and P. C. Kendall, {\it Variational techniques
including effective and weighted index methods}, in Progress in
Electromagnetic Research (EMW, Cambridge, Mass., 1995), Vol. 10, pp.
1-40.

\bibitem{Christodoulides03}
R. Morandotti, U. Peschel, J. S. Aitchison, H. S. Eisenberg, and Y.
Silberberg, Phys. Rev. Lett. {\bf 83}, 4756 (1999); T. Pertsch, P.
Dannberg, W. Elflein, A. Br\"{a}uer, and F. Lederer, Phys. Rev.
Lett. {\bf 83}, 4752 (1999); D. N. Christodoulides, F. Lederer, and
Y. Silberberg, Nature {\bf 424}, 817 (2003).

\bibitem{note}
The effective index $n_e(x)$ is defined from the eigenvalue equation
$\partial_y^2 G(x,y)+[n(x,y)/ \lambdabar]^2 G(x,y)=[n_e(x)/
\lambdabar]^2G(x,y)$ taking $x$ as a parameter (see Eq.(2) in Ref.
\cite{Benson}).

\bibitem{Trompeter06}
M. Ghulinyan, C. J. Oton, Z. Gaburro, L. Pavesi, C. Toninelli, and
D. S. Wiersma, Phys. Rev. Lett. {\bf 94}, 127401 (2005); H.
Trompeter, T. Pertsch, F. Lederer, D. Michaelis, U. Streppel, A.
Br\"{a}uer, and U. Peschel, Phys. Rev. Lett. {\bf 96}, 023901
(2006).

\bibitem{Lenz99} G. Lenz, I. Talanina, and C. M. de Sterke, Phys. Rev.
Lett. \textbf{83}, 963 (1999); N. Chiodo, G. Della Valle, R.
Osellame, S. Longhi, G. Cerullo, R. Ramponi, P. Laporta, and U.
Morgner, Opt. Lett. {\bf 31}, 1651 (2006).

\end{thebibliography}
\end{document}